\documentclass{article}
\usepackage[a4paper, total={7.2in, 9.5in}]{geometry}
\usepackage[utf8]{inputenc}
\usepackage{blindtext}
\usepackage{ragged2e}
\usepackage{multicol}
\usepackage[font=sf]{caption}
\usepackage{fancyhdr}
\usepackage[pdftex]{hyperref}
\usepackage[dvipsnames]{xcolor}
\usepackage[most]{tcolorbox}
\usepackage{tabularx}
\usepackage{enumitem,kantlipsum}
\usepackage[font=sf]{caption}
\usepackage[font=sf]{floatrow}

\hypersetup{
breaklinks=true,   
colorlinks=true,   
}
\definecolor{textgreen}{RGB}{33, 143, 38}
\definecolor{headergreen}{RGB}{19, 88, 97}

\begin{document}
\sffamily

\noindent{\Huge Sonification and Sound Design for Astronomy Research, Education and Public Engagement}

\vspace{0.2cm}{\large{\color{textgreen}\noindent A. Zanella$^{1,\star}$, C.M. Harrison$^{2,\dagger}$, S. Lenzi$^{3}$
J. Cooke$^{4,5}$, P. Damsma$^{6}$, S.W. Fleming$^{7}$}}\\
{\em $^{1}$INAF, Padova, Italy. $^{2}$Newcastle University, Newcastle, UK. $^{3}$Center for Design, CAMD, Northeastern University, Boston, USA. $^{4}$Centre for Astrophysics \& Supercomputing, Swinburne University of Technology, $^{5}$ARC Centre of Excellence for Gravitational Wave Discovery (OzGrav), Victoria, Australia. $^{6}$Sonokids Australia,  Robina, Queensland, Australia. $^{7}$Space Telescope Science Institute, Baltimore, USA}\\
\noindent Email: $^{\star}$\href{mailto:anita.zanella@inaf.it}{anita.zanella@inaf.it} $^{\dagger}$\href{mailto:christopher.harrison@newcastle.ac.uk}{christopher.harrison@newcastle.ac.uk}
 
\vspace{0.6cm}\noindent{\large Over the last ten years there has been a large increase in the number of projects using sound to represent astronomical data and concepts. Motivation for these projects includes the potential to enhance scientific discovery within complex datasets, by utilising the inherent multi-dimensionality of sound and the ability of our hearing to filter signals from noise. Other motivations include creating engaging multi-sensory resources, for education and public engagement, and making astronomy more accessible to people who are blind or have low vision, promoting their participation in science and related careers. We describe potential benefits of sound within these contexts and provide an overview of the nearly 100 sound-based astronomy projects that we identified. We discuss current limitations and challenges of the approaches taken. Finally, we suggest future directions to help realise the full potential of sound-based techniques in general and to widen their application within the astronomy community. }

\begin{multicols*}{3}
\noindent Visualisation, in the form of images, graphs and animations, is the standard approach used by astronomy researchers to inspect and present their data and analyses. Visualisation is also the primary method of communicating astronomical data and concepts to the general public and in educational settings. However, solely relying on visualisation has a number of limitations. For example, datasets are becoming extremely large and complex, often containing many dimensions. This makes it difficult to effectively and comprehensively display the data using standard visualisation techniques and it is often required to prioritise and filter the data to only visualise the information that is believed to be relevant. Importantly, this approach can involve making assumptions about the underlying information contained within the data, limiting the potential of making unexpected discoveries$^{1}$. Visualisation techniques can also be sub-optimal for representing non-stationary data, for fast identification, and for live data monitoring (e.g., for transients$^{2}$). 

\noindent Focusing on visualisation in both research and in educational contexts also naturally creates a barrier for people who are blind or vision impaired (BVI) to access astronomy knowledge, and related careers, starting from a very young age$^{3}$. As a consequence, there are only a handful of BVI professional astronomers and they face the additional burden of developing new tools and techniques to carry on their work$^{4,5}$. 

\begin{figure*}
\centerline{\includegraphics[width=0.9\textwidth]{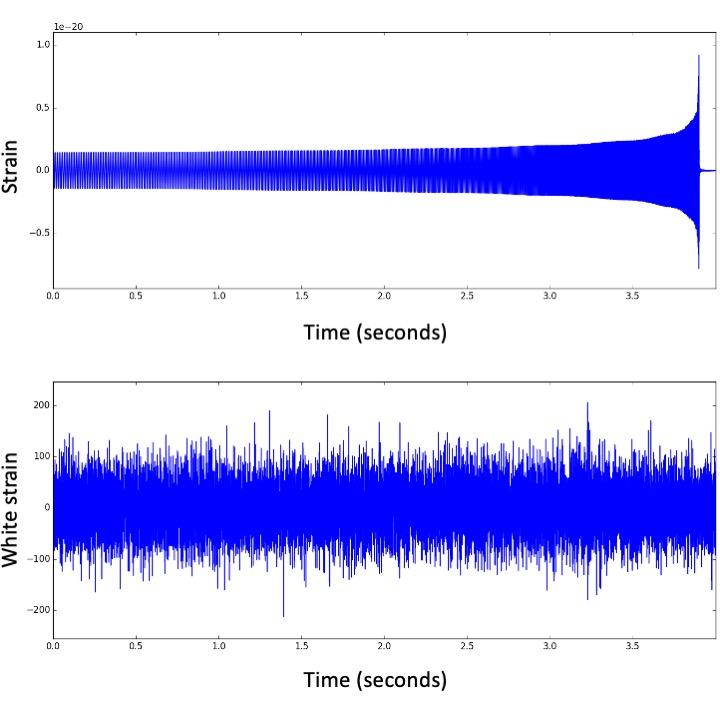}}
\RaggedRight{\footnotesize{\textbf{{\color{textgreen}Fig. 1.} Example of a simulated signal embedded in a dataset from the LIGO gravitational wave detector, plotted as gravitational wave strain versus time.} {\em Top panel:} Simulated noiseless signal. This is how the gravitational wave data would look like if there were no noise (i.e. no interference) and is represented with sound \href{https://youtu.be/6fIsogEDZEo}{at this link}. {\em Bottom panel:} The same simulated signal embedded in a noisy LIGO dataset. White strain is plotted against time. An audible version of this dataset can be listened to \href{https://www.youtube.com/watch?v=SYFN9i9tEmQ}{at this link}. The sonification has been done by the \href{https://blackholehunter.org/}{Black Hole Hunter} team as part of an online game. To make the game more engaging, in the noisy datasets (e.g., bottom panel) the signal has been shifted in time such that it could occur anywhere in the data except the first 1 and last 0.25 seconds. In this case the signal reaches its peak at $\sim$3.4 seconds. The change in scale for the y axis in the two panels is due to the whitening of the strain.\newline 
Credit: Connor McIsaac (University of Portsmouth) and Edward Fauchon-Jones (Cardiff University). }

}
\end{figure*}

\vspace{0.3cm}\noindent\textbf{Potential benefits of sound for astronomical research and communication}

\noindent Sound is inherently multi-dimensional because it is characterised by various parameters (e.g., pitch, volume, tempo, location in a stereo field, timbre). We can also perceive several sounds simultaneously, meaning that we can listen to different sonified streams in parallel and our hearing can focus on one of many audio streams, which is related to the so-called ‘cocktail party effect’$^{6}$. This allows the listener to detect weak information of interest amongst a background of noise. Therefore, there is a strong theoretical potential in using sound to help us intuitively and comprehensively explore large, noisy, complex and/or multi-dimensional datasets$^{7,8}$. This is a very relevant task in the era of astronomical ‘big data’. For example, in the case of a datacube (e.g., integral-field spectroscopy or submillimeter interferometry), sound could be used to represent one dimension (e.g., frequency) while visuals represent the two spatial dimensions. There is already initial evidence from small scale studies that using sound, in conjunction with visuals, to explore datasets can help make people more effective at gaining an initial overview of the data and/or for identifying signals and features in astronomical and space science data$^{1,9,10}$. 

\noindent Compared to vision, the ear is better at perceiving time-based information, patterns and transient changes$^{11,12}$ and does not require us to be oriented in the direction of the sound. Furthermore, hearing is always active, which makes it useful for monitoring alarms and continuous data streams, whereas an event may be missed with visual inspection due to blinking or looking away momentarily$^{13,14,15}$. An effective example of audible inspection of data used in the scientific context is the Geiger Counter, which clicks in response to invisible radiation levels. Therefore, sound has the potential to be a more effective alternative to visualisation for exploring time-series data and for live data monitoring of transient events whilst occupied with different tasks$^{1,2}$. Cooke et al.$^{2}$ plan to use their tool \href{https://www.jeffreyhannam.com/starsound}{StarSound} to audibly monitor fast transient data as part of their ‘Deeper, Wider, Faster’ programme. Hundreds of fast-evolving transient candidates are found every few minutes throughout an observing run via software and high-priority candidates need to be confirmed by human inspection within minutes to trigger follow-up observations before they fade.

\noindent An example of where a signal is more easily identified in a direct sound representation of noisy time-series data, compared to a direct visualisation of the same data, is presented in Figure 1. Here the signal of a simulated gravitational wave event is presented in the form of gravitational wave strain versus time$^{16}$. The simulated signal is then embedded into a noisy dataset from LIGO and turned into sound by creating a time-series of the data and mapping the strain into amplitude (the code is publicly available \href{https://github.com/icg-gravwaves/black_hole_hunter_sounds/blob/main/BHHData.py}{at this link}). The signal can easily be identified in the sound representation due to the different characteristic frequencies of the signal (increasing pitch) and the noise (steady low pitch), which our hearing can easily disentangle but our sight can not - at least when presented in this very basic format. We note that this particular example was made for the public engagement project \href{https://blackholehunter.org/game.html}{Black Hole Hunter}, communicating gravitational wave concepts to a non-specialised audience and it has not (yet) been demonstrated to be useful for research (see more discussion on the testing of the efficacy of sound-based approaches below).    

\noindent Another clear benefit of using sound, in combination with other multi-sensory learning schemes is that it allows BVI students to access science education at the same level as sighted students$^{17}$. Students with dyslexia or autism might also benefit from alternative learning modalities$^{18}$. Already, sound-based graphing tools have been shown to support BVI students’ independence in maths education and improve their engagement in the classroom$^{19}$. Therefore, it is clear that using sound-based representations for both education and public engagement has the potential to make astronomy more inclusive for BVI audiences and, indeed, for anybody who prefers more aurally-based learning methods$^{20,21}$.

\vspace{0.3cm}\noindent\textbf{Sound design and sonification}

\noindent Before exploring in more detail how sound-based techniques have been applied to astronomy applications, it is useful to briefly make two key definitions:

\noindent {\em Sound design} is the use of sound to make an intention audible or, in other words, to represent something other than itself such as an object, concept or system$^{22,23}$. In sound design, the sound might not directly connect to the data or phenomena itself. An example of this for communicating astronomy is using sounds of different sections of musical instruments to convey the different characteristics of rocky and gaseous planets in the Solar System$^{24}$.

\noindent {\em Sonification} is a technique for representing information and data using non-speech audio$^{25}$. Sonification can be considered a type of sound design where the sounds produced are tied to the data. One of the most common sonification methods is called parameter mapping$^{26}$. In this case, the sound characteristics (e.g., pitch, volume or timbre) are tied to the data. One example is the sonification of simulations of gravitational wave events performed by \href{https://blackholehunter.org/}{Black Hole Hunter} (Figure 1). For details on the different approaches of using sound to represent data and concepts in a wider range of contexts  we refer the reader to Herman et al.$^{7}$ and to the \href{https://icad.org/proceedings/}{ICAD proceedings archive}. 

\vspace{0.3cm}\noindent\textbf{Successful applications of using sound for astronomy}

\noindent There are some early examples where unexpected discoveries with astronomical origins were aided by listening.  In 1932 K. Jansky discovered that radio waves emitted by the centre of the Milky Way were responsible for some of the noise plaguing telephone communications at the time$^{27}$.  In 1964, A. Penzias and R. Wilson understood that the persistent noise disturbing the observations they were performing with an antenna were due to the cosmic microwave background radiation$^{28}$ (New York Times, \href{https://www.nytimes.com/1965/05/21/archives/signals-imply-a-big-bang-universe-signals-imply-a-big-bang-universe.html}{May 21, 1965}). Some sources report that listening to the data might have aided the discovery (e.g., \href{https://www.newscientist.com/article/dn25134-father-of-big-bang-carries-its-hiss-on-his-cellphone/}{this} interview with R. Wilson). The detection earned them the Nobel Prize in Physics.

\noindent There are later examples where conscious attempts to use sonification proved to be key to make discoveries in astronomy and space science$^{1,10}$.  For example, sonification techniques were used to identify the problem affecting the Voyager 2 space mission when it began its traversal of the rings of Saturn. The controllers were unable to pinpoint the problem using visual displays, which only showed noise. When the data were sonified and listened to, a "machine gun" sound was heard during a critical period, leading to the discovery that the problem was caused by high-speed collisions with electromagnetically charged micrometeoroids.$^{29}$ Landi et al.$^{30}$ used sonification to help identify the most promising carbon ionic ratio to measure the solar wind type. They mapped data into audio samples and after repeated close listening they could identify a pattern, a ``hum'' with a specific frequency (137.5 Hz) during the solar minimum of solar cycle, and overtones above this fundamental in the C$^{6+}$/C$^{4+}$ ratio that were absent in C$^{5+}$. Such a discovery was made specifically thanks to the adopted data sonification$^{31,32}$.

\noindent The first JavaScript-based sonification toolkit to be developed, which was xSonify$^{33,34,35}$, was used to test several sonification techniques on Hawkeye magnetic field and plasma measurements for solar magnetopause, bow shock and cusp crossings. Furthermore, this software was used as part of the work by a pioneer in the field of sonification at the professional level, Wanda Diaz-Merced, who is a blind astronomer and used sonification systematically for research. For example, Diaz-Merced et al.$^{36}$ used xSonify to search for the formation of irregularities in the ionosphere and to identify changes at the ionospheric recombination time.

\vspace{0.3cm}\noindent\textbf{Status of astronomy sonification and sound design projects}

\begin{figure*}
\centerline{\includegraphics[width=0.9\textwidth]{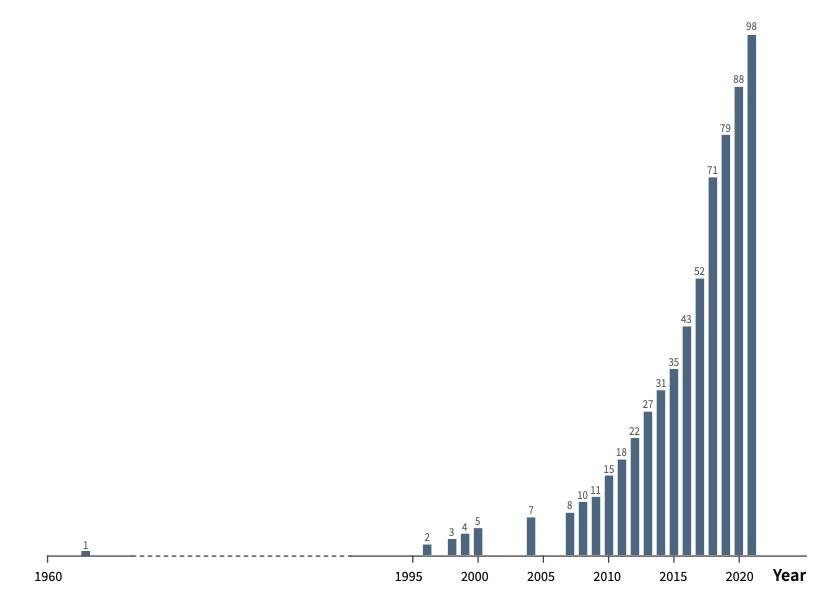}}\centerline{\includegraphics[width=0.9\textwidth]{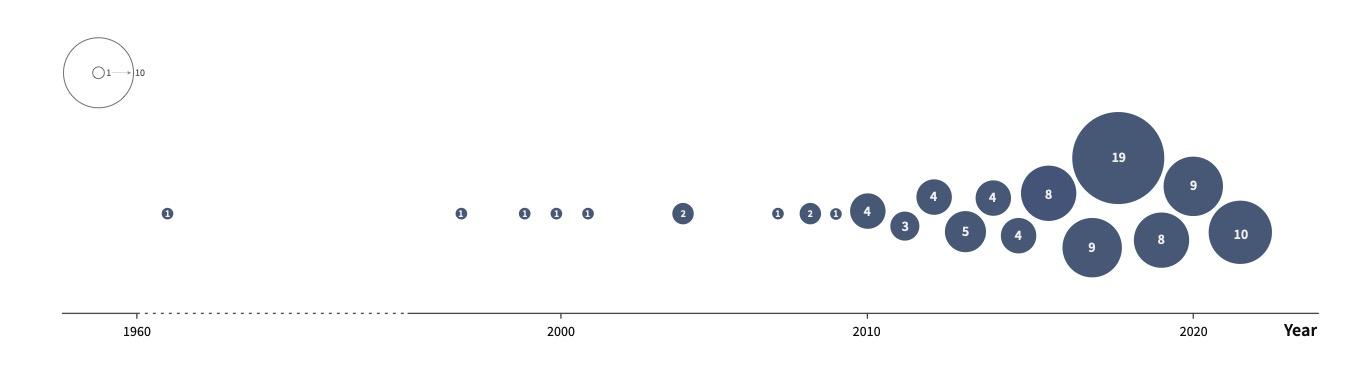}}
\RaggedRight{\footnotesize{\textbf{{\color{textgreen}Fig. 2.} Evolution in the number of sonification projects of space science and astronomical data based on project start date.} 
{\em Top panel:} Cumulative number of projects as a function of time. {\em Bottom panel:} Number of projects started each year. Data from this figure is also tabulated in Source Data Table 1. An audible version of these data can be listened to \href{https://www.youtube.com/watch?v=nRefoVtZYyo}{at this link}. The sonification has been created with the \href{https://www.jeffreyhannam.com/starsound}{StarSound tool}. The number of projects per year is mapped into pitch and the timbre of a bright acoustic piano was used. The cumulative number of projects has instead the sound of a Celesta. The pitches vary between 220 Hz and 550 Hz, with the lowest (highest) pitch representing the minimum (maximum) number of projects. The sounds are linearly spaced in time.\newline
Credit: Image by Yuan Hua. Sonification by Jeff Hannam.
}
}
\end{figure*}

\noindent For the rest of this article we focus on the conscious attempts to apply the techniques of sound design and sonification in an astronomy context. Through a community effort, we have collated 98 such applications, as of December 2021, into the repository Data Sonification Archive (\href{https://sonification.design/}{https://sonification.design/})$^{37}$, with an ``astronomy'' tag. This compilation is a combination of self-submissions to the archive, a targeted survey to people known to be interested in the topic, internet searches and discussions during a dedicated workshop$^{38}$. Although our compilation is likely not complete, we believe it is representative enough to draw basic conclusions about the trends and behaviours in this area. In the following we discuss the results obtained by analysing this data collection (Figure 2 and 3). The archive is open for new submissions and therefore can constantly be updated with new or historic examples. 

\noindent In Figure 2 we show the evolution in the number of known astronomy sound design and sonification projects through time. To our knowledge, the first documented and conscious attempts to sonify space science data are those made by Donald Gurnett, who from 1962 to 2012 used sound to analyse and convey information from different space missions, such as Cassini and Voyager. Sonification made possible the discovery of Saturnian lightnings, waves with frequency of few Hz to few kHz, and that of Saturn kilometric radiation, emitted by electrons in Saturn’s auroral zones$^{39}$. Since 1996, Figure 2 reveals a rise in the number of known sonification and sound design astronomy projects per year, with between 8 and 19 new projects launched each year since 2016.

\noindent In Figure 3 we show what the primary goals of sonification projects are: research, public engagement, inspiration for art, BVI accessibility and education. We also split projects based on their target audience: the general public and researchers (which we note includes students at the University level). Finally, Figure 3 shows whether sonification has been used in tandem with other media, such as data visualisation, graphic interfaces, videos, and haptic elements. This information is also presented in Source Data Table 2.

\noindent The majority of projects currently available have a primary goal of public engagement (about 36\%) or research (about 26\%). About 17\% have a primarily artistic purpose and only about 8\% are for education. Perhaps surprising, making astronomy BVI accessible was only listed as the primary goal for 13\% of the projects; however, a further 22\% of the projects mention accessibility as their secondary goal. This highlights that sonification is primarily being considered for a wide range of audiences, whilst increasing accessibility may be considered a natural additional outcome when multi-sensory data representations are adopted.

\noindent With the current set of projects, irrespective of their primary goal, most of the sonifications (79\%) are designed for the general public (which includes school pupils) as opposed to researchers (21\%). Interestingly, we found that even 30\% of the projects with research as their primary goal are using sonification primarily to help communicate the astronomical research to the general public (Figure 3 and Source Data Table 1). Although we do not have data on the number of researchers actually using sonfication for their work, these findings, along with anecdotal evidence suggests the number remains small at this time. Nonetheless, there is a clear increasing trend of using sonification for public engagement and education. 

\noindent Another finding from our surveys is that most of the projects use a multi-sensory approach (64\%) rather than sound only (Figure 3; Source Data Table 2). Most common is to combine sonification with visuals, with 62\% of projects combining sound with graphical visualisations, videos or a graphical user interface (GUI). Interestingly, most projects that have accessibility as their primary goal mix sonification with videos or a GUI. Possible reasons for this use of mixed media include the fact that the project aims at engaging both BVI and sighted users and therefore visuals are included (as recommended by Pérez-Montero$^{3}$) and/or the tool has been created by sighted developers who are used to working with a visual interface. We note that ensuring tool accessibility is critical and this has not always been successfully implemented during the development of sonification tools$^{40}$.

\noindent Only a minority of the projects (2\%) use haptic elements and these all have education as their primary goal (Figure 3). The haptic elements can be 3D models (e.g., Sense the Universe, Varano \& Zanella, submitted) or vibrations of the device (e.g., \href{https://www.a4bd.eu/}{A4BD} uses vibration to indicate the contours and shape of the image). However there are many more public engagement/accessibility astronomical projects available that use tactile supports, but they are not included in this review as they do not pair it with sound$^{41,42}$.

\noindent Finally, two thirds of the projects are interactive, as they allow a certain degree of user choice in the data-sound parameter mapping, the use of command line and/or graphical user interface, and/or the possibility to interact with the data or device.

\begin{figure*}
\centerline{\includegraphics[width=1.1\textwidth]{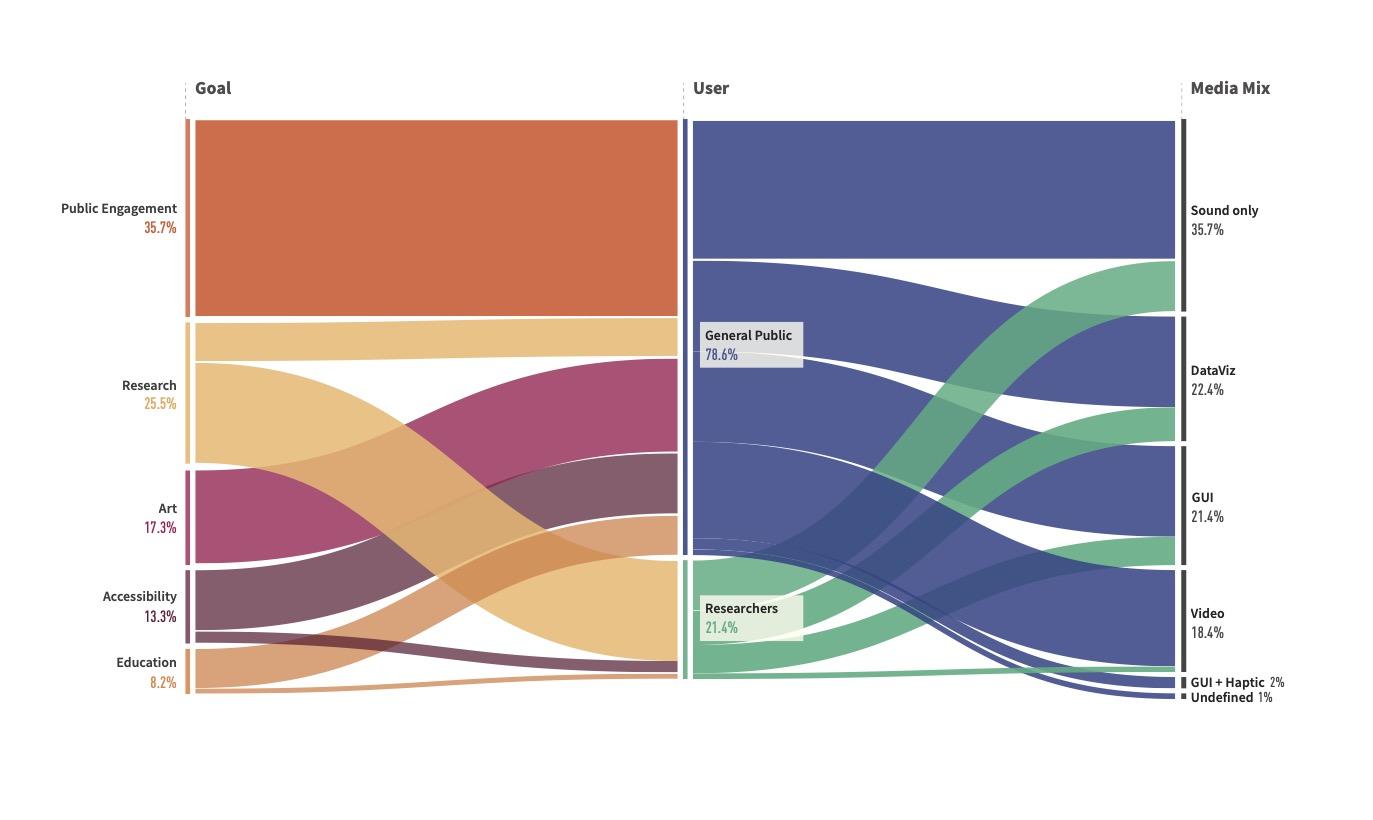}}
\RaggedRight{\footnotesize{\textbf{{\color{textgreen}Fig. 3.} Summary of goals, users, and media used by the 98 astronomy sonification applications reported in the \href{https://sonification.design/}{Data Sonification Archive} as of December 2021.} We group projects based on their specific goals (left-side vertical bars), users (central vertical bars), and media (right-side vertical bars). The alluvial diagram shows how specific goals are split over different users, and what type of media are used in combination with sound, for different users. For example, projects dedicated to Research (25.5\%), represented with a light orange colour, have Researchers (light green) as their main target group, and General Public (blue) as minority target. Projects dedicated to Education target mainly the General Public and, in a much smaller number, Researchers. Projects for the General Public (central axis, in blue) are primarily designed as sound-only experiences (35.7\%), but they also combine sound with Data Visualisation, Video, GUI and in a smaller percentage (2\%), with GUI and Haptic. The content of this Figure is also tabulated in Source Data Table 2. \newline 
Credit: Image by Yuan Hua. 
}
}
\end{figure*}

\vspace{0.2cm}\noindent{\em Common themes and trends of astronomy sonification and sound design}

\noindent Currently, the approach taken for the sonification or sound design for astronomy applications is not standardised, which is one of the major limitations preventing sonification from becoming a mainstream tool. However, commonalities in the way different groups and projects sonify data can already be identified. For example, when the project or tool addresses the public, it focuses more on inspiring the audience and conveying a single message rather than being closely related to the underlying data. The opposite is true for projects meant for researchers. For public audiences, the sound is carefully designed and there is extra attention to the pleasantness of the sonification: aesthetics is more important than accurate parameter mapping in these cases (e.g., in planetarium shows$^{24,43}$). This is even more extreme when data are used primarily for artistic inspiration, where the scientific content of the representation is minimal and aesthetics is the main driver (e.g., \href{https://www.agalaxyofsuns.net/}{A  Galaxy of Suns}). This approach is similar to the one used to visually represent the data: beautiful pictures are shown to the public, whereas raw data are used for research purposes$^{44}$.

\noindent Most current sonification projects designed for astronomy research sonify one-dimensional (e.g., spectra) or time series data (e.g., light curves) and two-dimensional images or graphs. The sonification of three-dimensional and in general multi-dimensional datasets is instead less advanced and only a few tools are currently available. Most one-dimensional sonifications scroll through the data, playing a single data point at a time (e.g., \href{https://astronify.readthedocs.io/en/latest/}{Astronify}), whereas two-dimensional sonifications allow the user to explore the image or graphs with their fingers/cursor (e.g., VoxMagellan) or scan through the image hearing one whole dimension simultaneously or multiple parameters at once in either multiple tones or the timbre of the tone (e.g., \href{https://afterglow.skynetjuniorscholars.org/}{Afterglow Access} and \href{https://astreos.space/}{AstreOS}). 

\noindent The majority of astronomy sonification projects link data to sound characteristics by using the parameter mapping technique. Pitch is the most commonly used auditory dimension, explained by the fact that it is known to be the most prominent sound attribute$^{45}$ and that humans remember pitch relationships (e.g., melodies) better than loudness relationship or timbre relationship$^{46}$. Pitch is generally associated with the dependent variable data dimension in astronomy applications as it is often associated with brightness. Sound spatialization is also used and generally associated with sky coordinates. Some projects have begun to exploit timbre for classification purposes (e.g., \href{https://strauss.readthedocs.io/en/latest/}{STRAUSS} for the classification of galaxy spectra) and loudness has been used at times to represent distance (Varano \& Zanella, submitted). Surprisingly, duration has rarely been used, despite the good capability of humans to discriminate against sound duration. Due to the temporal nature of sound, this could be explained by a sort of bias making one of the most obvious mappings implicit$^{45}$. Uncertainties are generally not included in the sonic representation, although some attempts exist (e.g., \href{https://www.jeffreyhannam.com/starsound}{StarSound}, and \href{http://sion.frm.utn.edu.ar/sonoUno/}{SonoUno}). Despite all of this work, there have been little-to-no attempts during the development of these tools to do extensive testing to establish the best approach to sound mapping (see also next Section).

\vspace{0.3cm}\noindent\textbf{Challenges and proposed solutions}
While the number of sound design and sonification projects of astronomical data has steadily increased (Figure 2), sound design and sonification is not yet incorporated into mainstream research tools, and remains a niche approach for education and communication. Possible reasons preventing sonification from becoming mainstream are: (1) a lack of training and familiarity in sonification and; (2) a lack of standardisation, evaluation and dissemination. We explore these in more detail below.  

\noindent{\em Training and familiarity}

\noindent Using the approach of sonification to explore data requires the listener to learn how to convert the characteristics of the sound into the properties of the data, with so-called ‘reduced listening’$^{47,48}$. Our current education and research methods focus on visualisation, and we are not trained to listen attentively for the purpose of gaining and analysing complex information, a required skill to effectively use sonification$^{49}$. Indeed, a recent online survey showed that astronomy and data analysis experts (i.e., those with relevant PhDs and careers) performed no better than non-experts at identifying mock signals of planet transits in sonified versions of noisy light curves; however, the experts performed significantly better than the non experts when using standard visualisations of the same data sets (Tucker Brown et al., in prep). This study suggested that sonification could be used to effectively identify signals (at least for high signal to noise ratios); however, just like for interpreting graphical representations, training and familiarity are crucial being effective and efficient at interpreting noisy data.   

\noindent Introducing attentive listening and sonification tools into the mainstream education curriculum, in tandem with data visualisation, would be a first step towards training the next generation to untap the potential of using sound for analysis and interpretation$^{4,17,50}$. Sonification would then also be a powerful tool to adopt to ensure the accessibility of astronomy from schools through to the academic environment. Towards this goal, several promising calculator and graphical applications that include speech outputs and sonification have been developed in the last years, such as \href{https://www.desmos.com/calculator}{Desmos graphing calculator}, \href{https://support.sas.com/software/products/graphics-accelerator/}{SAS Graphic Accelerator}, and Stocks (pre-installed app on Apple’s Iphone and Ipad). Making Desmos the default calculator for mainstream High School online exams would be key to allow all students to use the same calculators, independently on their level of vision and preferred learning style.

\noindent{\em Standardisation, Evaluation and Dissemination}

\noindent There is a long history of visual presentation of data that allowed us to develop sophisticated graphics and visualisations with now widely-accepted standard approaches$^{20}$. In contrast, we have yet to converge on a set of standard approaches to represent data through sound$^{1}$. Adopting universally accepted standards is key to making sonification generally understandable and mainstream. For increasing accessibility, it would also mean that BVI users did not need to create and define new approaches from scratch. 

\noindent The lack of systematic published evaluation of the usefulness and effectiveness of sonification in general, prevents its widespread adoption, especially for research purposes$^{45}$. Diaz-Merced$^{1}$ came to a similar conclusion, specifically for astronomy applications, reporting that most publications on sonification of numerical data focus on sonification techniques rather than evaluating their usability. 

\noindent During a multidisciplinary workshop$^{38}$ we established that various sound-based astronomy projects have gathered anecdotal or informal evidence on the efficacy of sonification (e.g., \href{https://astreos.space/}{AstreOS}, \href{https://www.a4bd.eu/}{A4BD}). These have led to positive feedback from users, in particular for inspiring BVI children and adults. However, only a few projects have, or are in the process of, publishing quantitative evaluation of their sonifications$^{43}$ (Varano \& Zanella, submitted, Tucker Brown et al., in prep). Nonetheless, these studies do not explore multiple sonification approaches, for example changing the parameter mapping, to establish the most effective approach in different applications. This would be a first step towards establishing standards and preventing different groups by repeating the creation of similar sonification tools that might not use the most optimal approaches. 

\noindent Quantitatively demonstrating the usefulness of sonification and sound design would help convince sceptical academics and educators to consider these approaches and, in turn, to convince funding agencies to support this line of research. To make sonification a more robust and widespread representation method of astronomical datasets we urge the community to carefully define their goals from the beginning of the design phase, as well as defining and carrying on a rigorous evaluation plan, possibly following the guidelines highlighted by Lenzi et al.$^{51}$, Giordano et al.$^{52}$ and Misdariis et al. (in prep).

\noindent We note that one challenge to the creation of universally accepted standards are cultural differences, such as the use of different musical scales and harmonies. However to guide choices it is possible to draw upon some everyday experiences which may cross cultural barriers, such as the fact that heavier objects sound louder when dropped; the damping of sound waves in the transmission medium decreases the sound intensity as the distance of the sound source increases; the frequency of a sound wave changes due to the Doppler effect, depending on the relative position between the listener and the sound source$^{8}$. 

\noindent Having the community systematically publishing peer-reviewed articles on their findings related to sonification of astronomical data, having conferences that encourage speakers to use sonifications when they present graphs, and having academic journals sonifying graphs and visual contents would help make sonification more mainstream in the astronomical academic environment.

\vspace{0.3cm}\noindent\textbf{Concluding remarks}

\noindent The number of sonification projects in astronomy is increasing rapidly and there are examples where sonification of astronomical data has been successfully used to engage the general public with astronomy, for example in planetarium shows$^{53,54}$ and for accessibility purposes$^{17,24,43,55}$. There are also a handful of discoveries that have been aided by listening to the data$^{1,30}$. These examples, and other cases, demonstrate a clear potential for using sound design and sonification in astronomy research, outreach, and education, with benefits for both science research and accessibility. 

\noindent The next steps needed to make sonification mainstream in astronomy include extensively evaluating its efficacy and looking for standardised solutions for quantitative data analysis (e.g., for uncertainty representation). Similarly, it is important  to understand how to standardise sonification methods to efficiently present a quick overview of the  astronomical datasets, before digging into more details (i.e., the sonic equivalent of a thumbnail image). New solutions could also be investigated, such as how to use sound filters to better distinguish the signal from the noise, how to properly represent axes labels and tick marks, how to provide the user with flexible control over data dimension and sound parameters, and how to facilitate the exchange of data formats and display systems$^{20}$. Cross-disciplinary collaborations and proper dissemination will be key to address these challenges and advance in the field.

\vspace{0.5cm}\noindent{\color{textgreen}References}
\vspace{-0.15cm}

\begin{enumerate}[leftmargin=*]
  \setlength{\itemsep}{1pt}
  \setlength{\parskip}{0pt}
  \setlength{\parsep}{0pt}
\footnotesize{
\item Diaz-Merced, W. L. ``Sound for the Exploration of Space Physics Data'' PhD thesis University of Glasgow (2013)

\item Cooke, J. et al., ``Exploring Data Sonification to Enable, Enhance, and Accelerate the Analysis of Big, Noisy, and Multi-Dimensional Data'', Proceedings of the International Astronomical Union, IAU Symposium, 339, 251 (2019)

\item Pérez-Montero, E. Towards a more inclusive outreach. Nat. Astron. 3, 114-115 (2019).

\item Bourne, R. L. A., Flaxman, S. R., Braithwaite, T. et al., ``Magnitude, temporal trend, and projections of the global prevalence of blindness and distance and near vision impairment: a systematic review and meta-data analysis'', The Lancet, 5, 9, 888 (2017)

\item Noel-Storr \& Willebrands, ``Accessibility in Astronomy for the Visually Impaired'', Nature Astronomy, DOI: 10.1038/s41550-022-01691-2

\item Bronkhorst, A. W., ``The cocktail party phenomenon: A review of research on speech intelligibility in multiple-talker conditions,'' Acta Acustica united with Acustica, vol. 86, pp. 117–128 (2000)

\item Hermann T, Hunt T, Neuhoff JG, ``The sonification handbook'', Logos Publishing House, Berlin, Germany (2011)

\item Sawe, N., Chafe, C., and Trevinoo, J., ``Using data sonification to overcome science literacy, numeracy, and visualization barriers in science communication,'' Frontiers in Communication, vol. 5, p. 46, (2020)

\item Tutchton, R. M., Wood, M. A., Still, M. D., Howell, S. B., Cannizzo, J. K., Smale, A. P., ``Sonification of Kepler field SU UMA cataclysmic variable stars V344 Lyr and V1504 Cyg'', Journal of Southeastern Association for Research in Astronomy, 6, 21 (2012)

\item Diaz-Merced, W. L. ``Sonification of Astronomical Data'', New Horizons in Time-Domain Astronomy Proceedings IAU Symposium No. 285, R.E.M. Griffin, R. J. Hanish, R. Seaman eds. (2011)

\item Guttman, S. E., Gilroy, L. A., and Blake, R., ``Hearing what the eyes see: auditory encoding of visual temporal sequences,'' Psychological science, vol. 16, no. 3, pp. 228–235 (2005)

\item Walker, B. N. and Nees, M. A., ``Theory of Sonification'', In: Hermann T, Hunt T, Neuhoff JG (eds) The sonification handbook. Logos Publishing House, Berlin, Germany, pp. 9-40 (2011)

\item Tzelgov, J., Srebro, R., Henik, A., \& Kushelevsky, A. ``Radiation detection by ear and by
eye.'' Human Factors 29(1), 87-98 (1987)

\item Pauletto, S. \& Hunt, A., ``A comparison of audio and visual analysis of complex time-series data'', Proceedings of ICAD 05 (2004) 

\item Lunn, P. and Hunt, A. ``Listening to the invisible: Sonification as a tool for astronomical
discovery'' In: Making visible the invisible: art, design and science in data visualisation, 10th-11th
March 2011, Huddersfield, UK (2011)

\item Abbott, B. P. et al., ``Observations of gravitational waves from a binary black hole merger'', Physical Review Letters, 116, 6, (2016)

\item Ediyanto and N Kawai, ``Science Learning for Students with Visually impaired: a Literature Review'' J. Phys.: Conf. Ser. 1227 012035 (2019)

\item Rose, D. \& Meyer, A. ``Teaching Every Student in the Digital Age: Universal Design for Learning'', Virginia: Association for Supervision and Curriculum Development (2002)

\item Tomlinson, B. J. et al. ``Exploring auditory graphing software in the classroom: The effect of auditory graphs on the classroom environment'', ACM Trans. Access. Comput. 9, 1, Article 3 ( 2016)

\item Kramer, Gregory; Walker, Bruce; Bonebright, Terri; Cook, Perry; Flowers, John H.; Miner, Nadine; and Neuhoff, John, ``Sonification Report: Status of the Field and Research Agenda'' Faculty Publications, Department of Psychology 444 (2010)

\item National Academies of Sciences, Engineering, and Medicine ``Pathways to Discovery in Astronomy and Astrophysics for the 2020s'', Washington, DC: The National Academies Press (2021)

\item Susini P., Houix P., Misdariis N., ``Sound design: an applied, experimental framework to study the perception of everyday sounds'', The New Soundtrack, 4, 103-121 (2014)

\item Pauletto S., ``Special Issue: Perspectives on Sound Design'', The New Soundtrack, 4, v-vi (2014)

\item Harrison, C. et al., ``Audio Universe Tour of the Solar System: using sound to make the universe more accessible'', Astronomy \& Geophysics (2022), 63, Issue 2, p. 2.38-2.40

\item Kramer, G., Walker, B. N., Bonebright, T., Cook, P., Flowers, J., Miner, N., et al. ``The Sonification Report: Status of the Field and Research Agenda.'' Report prepared for the National Science Foundation by members of the International Community for Auditory Display. Santa
Fe, NM: International Community for Auditory Display ICAD (1999)

\item Grond, F. and Berger, J. ``Parameter mapping sonification'' In: Hermann T, Hunt T, Neuhoff JG (eds) The sonification handbook. Logos Publishing House, Berlin, Germany, pp 399–427 (2011)

\item Kellermann K.I., Bouton E.N., Brandt S.S., ``A New Window on the Universe'', In: Open Skies. Historical \& Cultural Astronomy. Springer, Cham. (2020)

\item Penzias A. A., Wilson R. W., ``A measurement of excess antenna temperature at 4080 Mc/s'', Astrophys. J., 142, 419 (1965)

\item Scarf F.L. ``Voyager 2 Plasma Wave Observations at Saturn.'' Science, 215, 
7-594 (1982)

\item Landi E. et al., ``Carbon ionization stages as a diagnostic of the solar wind'', The Astrophysical Journal, 744, 100 (2012)

\item Alexander, R. L., Gilbert, J., Landi, E., et al. 2011, Proc. 17th Int. Conf. Aud. Disp. (ICAD 2011)

\item Hadhazy, A., ``Heavenly Sounds: Hearing Astronomical Data Can Lead to Scientific Insights'', The Sciences (2014), link: \href{https://www.scientificamerican.com/article/heavenly-sounds-hearing-astronomical-data-can-lead-to-scientific-insights/}{Scientific American Online}. 

\item Candey, R. M., Kessel, R. L., Plue, J. R ``Web-based sonification of space science data.'' SIGGRAPH98 Conf. Abstracts and Applications, 166 (1998)

\item Candey, R. M., A.M. Schertenleib, and W.L Diaz Merced. ``Sonification Prototype for Space Physics.'' AGU, 52 (2005)

\item Candey, R.M., Schertenleib, A. M., and Diaz Merced, W.L., ``xSonify Sonification Tool For Space Physics'' in Proc. of the ICAD, 12th International Conference on Auditory Display, London, UK (2006)

\item Diaz-Merced, W. L., Candey, R. M., Mannone, J. C., David, F., Rodriguez, E. ``Sonification for the analysis of plasma bubbles at 21 MHz'', Sun and Geosphere, 3, 42 (2008)

\item Lenzi S., Ciuccarelli P., Liu H., Hua Y. (2020). Data Sonification Archive. \href{http://www.sonification.design}{http://www.sonification.design}. Last accessed: 12/2021

\item Harrison, C., Zanella, A., Bonne, N., Meredith, K., and Misdariis, N. ``Audible Universe'', Nat. Astron., 6, 22 - 23, (2021).

\item Gurnett, D.A. Cassini Encounters Saturn's Bow Shock. 2004 06 \href{http://www-pw.physics.uiowa.edu/space-audio/cassini/bow-shock/}{http://www-pw.physics.uiowa.edu/space-audio/cassini/bow-shock/} 

\item Garcia, B., Diaz-Merced, W. L., Casado, J., Cancio, A. ``Evolving from xSonify: a new digital platform for sonorization'', EPJ Web of Conferences, 200, 01013 (2019)

\item Bonne, N., Gupta, J., Krawczyk, C. and Masters, K., ``Tactile Universe makes outreach feel good'', Astronomy \& Geophysics 59a, p. 1.30 (2018)

\item Paredes-Sabando, P. and FuentesMunoz, C., ``Dedoscopio Project: Making Astronomy Accessible to Blind and Visually Impaired (BVI) Communities Across Chile'', CAPjournal 29, p. 27 (2021)

\item Elmquist, E., Ejdbo, M., Bock, A., Rönnberg, N., ``Openspace sonification: complementing visualization of the solar system with sound'', Proceedings of the 26th International conference on auditory display ICAD (2021)

\item Barrass S., ``The aesthetic turn in sonification towards a social and cultural medium'', AI and Society, 27, 177 (2012)

\item Dubus G, Bresin R. ``A systematic review of mapping strategies for the sonification of physical quantities.'' PLoS One. 8, 12 (2013)

\item McDermott, J. H., Lehr, A. J., \& Oxenham, A. J. ``Is relative pitch specific to pitch?'', Psychological Science, 19, 1263 (2008)

\item Chion, M., ``The Audiovision: Sound on Screen'', New York, U.S.: Columbia University Press (1994)

\item Gaver, W, ``The SonicFinder: an interface that uses auditory icons'', Hum.-Comput. Interact 4, 1, 67 (1989)

\item Scaletti, C. "Why sonification is a joke. Keynote address delivered at the 23rd International Conference on Auditory Display", ed. University Park, PA., 2017, p. \href{https://www.youtube.com/watch?v=T0qdKXwRsyM}{YouTube: T0qdKXwRsyM}

\item Mozolic, J. L., Hugenschmidt, C. E., Peiffer, A. M., Laurienti, P. J., ``Modality-specific selective attention attenuates multisensory integration'', Exp. Brain. Res., 184, 39 (2007)

\item Lenzi S, Terenghi G, Moreno Ferdnadez-de-Leceta, A, ``A design-driven sonification process for supporting expert users in real-time anomaly detection: Towards applied guidelines.'' Endorsed Transaction on Creative Technology: 7(23). Special Issue on Sound, Image and Interactivity (2020)

\item Giordano, B. L., Susini, P., Bresin, R., ``Perceptual evaluation of sound-producing objects'', in K. Franinovic and S. Serafin (Eds.), ``Sonic interaction design''. Boston, MA: MIT Press, pp. 151-197 (2013)

\item Quinton, M. \& McGregor, I. \& Benyon, D. ``Sonifying the Solar System''. 28-35, ICAD proceedings (2016)

\item Tomlinson, B. J. et al. ``Solar System Sonification: Exploring Earth and its Neighbors Through Sound.'', ICAD proceedings (2017) 

\item Bieryla, A. et al., ‘LightSound: The Sound of An Eclipse’, CAPjournal, no. 28, 3 (2020)
}
\end{enumerate}

\vspace{0.3cm}\noindent{\color{textgreen}Acknowledgements}

\noindent{\small 
\noindent The authors are grateful to the Lorentz Centre for supporting the organisation of the ``Audible Universe'' workshop in September 2021 and to the workshop participants for valuable and insightful discussions. We also thank members of the Sonification World Chat and Kala Perkins, for helping us collate information about current astronomy sonification and sound design projects. We thank I. Harry, C. McIsaac, and S. Fairhurst for providing information about their Black Hole Hunter project and the possibility to include the data in Figure 1. We are grateful to Y. Hua and P. Ciuccarelli for their help in creating Figure 2 and 3. We are grateful to Jeff Hannam for his help in creating the sonification of Figure 2. We thank the referee, Matt Russo, for their constructive comments and ideas.
}

\vspace{0.3cm}\noindent{\color{textgreen}Author Contributions}

\noindent{\small 
\noindent A.Z. and C.M.H. led the writing and the researching of current astronomy sonification and sound design projects, S.L. led the preparation of Figures 2 and 3, all co-authors participated in the discussion of the content and provided comments to the manuscript.
}

\vspace{0.3cm}\noindent{\color{textgreen}Competing interests}

\noindent{\small The authors declare no competing interests.}

\end{multicols*}

\vspace{0.3cm}\noindent\textbf{\color{textgreen}Source Data}

\begin{table}[!h]
    \centering
    \begin{tabular}{ccc}
\hline
{\bf Year} & {\bf Number of projects} & {\bf Cumulative number of projects} \\
\hline
1962 &
1 &
1 \\
1997 &
1 &
2 \\
1999 &
1 &
3 \\
2000 &
1 &
4 \\
2001 &
1 &
5 \\
2004 &
2 &
7 \\
2007 &
1 &
8 \\
2008 &
2 &
10 \\
2009 &
1 &
11 \\
2010 &
4 &
15 \\
2011 &
3 &
18 \\
2012 &
4 &
22 \\
2013 &
5 &
27 \\
2014 &
4 &
31 \\
2015 &
4 &
35 \\
2016 &
8 &
43 \\
2017 &
9 &
52 \\
2018 &
19 &
71 \\
2019 &
8 &
79 \\
2020 &
9 &
88 \\
2021 &
10 &
98 \\
\hline
    \end{tabular}
    \caption{Evolution in the number of sonification projects of space science and astronomical data. This Table summarises the information shown in Figure 2. Columns: (1) Starting year of the project; (2) Number of projects started that year; (3) Cumulative number of projects.}
    \label{tab:project_timeline}
\end{table}

\begin{table}[]
    \centering
    \begin{tabular}{ccc}
\hline
\textbf{Primary goal} & \textbf{Primary user} & \textbf{Example} \\
\hline\vspace{0.1cm}
Public engagement (35.7\%) & General Public & \href{https://archive.org/details/PlanethesizerWindows}{Planethesizer} \vspace{0.2cm}\\
 Research (25.5\%) & General Public & \href{https://soniverse.space/galaxy-player/}{Galaxy Player} \\
& Researchers & \href{https://www.youtube.com/watch?v=gQdvv-gCa3Y}{VoxMagellan} \vspace{0.2cm}\\
 Art (17.3\%) & General Public & \href{https://goo.gl/EUm3ZB}{Kosmophone} \vspace{0.2cm} \\
Accessibility (13.3\%) & General Public & \href{https://www.audiouniverse.org}{Audio Universe: Tour of the Solar System} \\
& Researchers & \href{http://sion.frm.utn.edu.ar/sonoUno/}{SonoUno} \vspace{0.2cm} \\
Education (8.2\%) & General Public & \href{https://www.sonoplanet.com/}{Sonoplanet} \\
& Researchers & \href{http://www.jb.man.ac.uk/~pulsar/Education/Sounds/sounds.html}{The Sound of Pulsars} \vspace{0.5cm}\\
\hline
\textbf{User} &
\textbf{Media additional to sound} &
\textbf{Example} \\
\hline
General Public (78.6\%) & Sound only & \href{https://www.almaobservatory.org/en/alma-sounds/}{ALMA Sounds} \\
& Data Visualisation & \href{https://www.ithaca.edu/faculty/lkeller/research/spectral-sonification}{SonifySpec} \\
& GUI & \href{https://www.system-sounds.com/saturn-harp/}{Saturn Harp}\\
& Video & \href{https://www.system-sounds.com/trappist-sounds/}{TRAPPIST-1} \\
& Haptic+GUI & \href{http://www.eclipsesoundscapes.org/}{Eclipse Soundscape} \\
& Undefined & \href{https://exoplanets.nasa.gov/exoplanet-watch/about-exoplanet-watch/team/}{Exoplanet Watch}\vspace{0.2cm} \\

Researchers (21.4\%)& Sound only & \href{https://www.lisamission.org/multimedia/audio}{LISA consortium: gravitational waves}\\
& Data Visualisation& \href{https://astrosom.com/Apr2018.php}{The Sound of a Fast Radio Burst}\\
& GUI & \href{https://www.jeffreyhannam.com/starsound}{StarSound}\\
& Video & \href{https://www.youtube.com/watch?v=OOszY79xSkQ&feature=emb_title}{Solar Wind Audification}\\
\hline
    \end{tabular}
    \caption{Summary of goals, users, and media used by the astronomy sonification applications reported in the \href{https://sonification.design/}{Data Sonification Archive} as of December 2021. This Table summarises the information shown in Figure 3. }
    \label{tab:my_label}
\end{table}

\end{document}